\documentclass[12pt,amsart]{iopart}
\usepackage{fullpage,graphicx,epsf}

\usepackage{iopams}%
\usepackage{amssymb}
\usepackage{amscd}
\usepackage{color}%
\usepackage{graphicx}
\usepackage{epsfig}
\usepackage[english]{babel}
\usepackage{amsfonts}
\newcommand{\be}{\begin{equation}}
\newcommand{\ee}{\end{equation}}
\newcommand{\bea}{\begin{eqnarray}}
\newcommand{\eea}{\end{eqnarray}}
\newcommand{\bef}{\begin{figure}[h!]}
\newcommand{\eef}{\end{figure}}
\newcommand{\la}{\langle}
\newcommand{\ra}{\rangle}
\begin{document}
\title{Robustness of spontaneous symmetry breaking in a bridge model}
\author{Shamik Gupta$^1$, David Mukamel$^1$ and Gunter M. Sch\"{u}tz$^2$}
\address{%
$^1$Department of Physics of Complex Systems, Weizmann Institute of Science, Rehovot 76100, Israel\\
$^2$Institut f\"{u}r Festk\"{o}rperforschung, Forschungzentrum J\"{u}lich, D-52425 J\"{u}lich, Germany
}%
\ead{shamik.gupta@weizmann.ac.il, david.mukamel@weizmann.ac.il, g.schuetz@fz-juelich.de}
\begin{abstract}
A simple two-species asymmetric exclusion model in one dimension with bulk and boundary exchanges of particles is investigated for the existence of spontaneous symmetry breaking. The model is a generalization of the `bridge' model for which earlier studies have confirmed the existence of symmetry-broken phases, and the motivation here is to check the robustness of the observed symmetry breaking with respect to additional dynamical moves, in particular, the boundary exchange of the two species of particles. Our analysis, based on general considerations, mean-field approximation and numerical simulations, shows that the symmetry breaking in the bridge model is sustained for a range of values of the boundary exchange rate. Moreover, the mechanism through which symmetry is broken is similar to that in the bridge model. Our analysis allows us to plot the complete phase diagram of the model, demarcating regions of symmetric and symmetry-broken phases.    
\end{abstract}
\pacs{02.50.Ey, 47.11.Qr, 05.70.Ln, 73.22.Gk}
\date{\today}
\maketitle
\section{Introduction}
One-dimensional systems with short range interactions and a finite state space for the local variable, when driven far from equilibrium, often exhibit phenomena which are unexpected in equilibrium \cite{mukamel-rev, schutz-twospecies}. An example is that of spontaneous symmetry breaking which was observed in a driven system of two species of particles \cite{mukamel-ssb1, mukamel-ssb2}. In this so-called bridge model, one considers a one-dimensional lattice with sites occupied by two species of hard core particles, referred to below as the positive and the negative particles. The positive particles move stochastically to the right, while the negative particles move stochastically to the left. At the left boundary site, positive particles may enter the lattice and negative particles may leave; at the right boundary site, negative particles may enter the lattice and positive particles may leave. The dynamics is symmetric with respect to ``charge'' conjugation combined with space inversion. 

At long times, the system reaches a nonequilibrium stationary state with non-zero particle currents. For low extraction rate of particles, the system is typically loaded with a majority species of particles, say, the positive particles, with larger current and higher bulk density than the minority species (the negative particles).  As the system evolves, it flips between this state and the one in which the current and the density inequalities are reversed. General considerations and Monte Carlo simulations showed that the average time between flips grows exponentially with the system size, leading to spontaneous symmetry breaking (SSB) in the thermodynamic limit \cite{mukamel-ssb1}. The occurrence of SSB for the bridge model was demonstrated within mean-field approximation \cite{mukamel-ssb1, mukamel-ssb2} and also by rigorous results under specific conditions \cite{mukamel-ssb-rw, schutz-ssb1, schutz-ssb2}. Over the years, such symmetry breaking has also been observed in many variants of the bridge model \cite{popkov-ssb, levine-ssb, lipowsky-ssb, kolomeisky,jiang-1,jiang-2, jiang-3,bridge-junction}. It may be noted that, for a periodic system, the particle number is conserved and hence, SSB cannot occur. Therefore, SSB in the bridge model may be understood as a boundary-induced critical phenomenon.

In this paper, we study the robustness of SSB in the bridge model with respect to additional dynamical moves at the boundary. It is known that far from equilibrium, boundary-induced critically phenomena generally depend sensitively on microscopic details of the boundary processes \cite{asep-rev}. It is thus of interest to test the effect of more general boundary processes on the stationary state. To this end, we generalize the bridge model to allow for boundary exchange of particles of the two species. Our model is defined as follows.

We consider a one-dimensional lattice of $N$ sites. Each site $i$ of the lattice is occupied by either a positive ($+$) or a negative ($-$) particle, or is left vacant, denoted by $0$ (a ``hole''). The system evolves according to a stochastic Markovian dynamics. In an infinitesimal time $dt$, the following exchanges may take place at a pair of nearest-neighbor sites $(i, i+1)$ in the bulk ($1\le i \le N-1$): 
\bea
(+)_{i}(0)_{i+1} &\rightarrow& (0)_{i}(+)_{i+1} \mathrm{~with ~probability~} dt, \nonumber \\
(0)_{i}(-)_{i+1} &\rightarrow& (-)_{i}(0)_{i+1} \mathrm{~with ~probability~} dt, \label{updatebulk} \\
(+)_{i}(-)_{i+1} &\rightarrow& (-)_{i}(+)_{i+1} \mathrm{~with ~probability~} qdt. \nonumber 
\eea
At the boundaries, particles enter or leave the lattice. During an infinitesimal time $dt$, the following events may take place at the left boundary site ($i=1$):
\bea
(0)_{1} &\rightarrow& (+)_{1} \mathrm{~with ~probability~} \alpha dt, \nonumber \\
(-)_{1} &\rightarrow& (0)_{1} \mathrm{~with ~probability~} \beta dt, \label{updateleft} \\
(-)_{1} &\rightarrow& (+)_{1} \mathrm{~with ~probability~} \gamma dt, \nonumber 
\eea
while the following events may take place at the right boundary site ($i=N$):
\bea
(0)_{N} &\rightarrow& (-)_{N} \mathrm{~with ~probability~} \alpha dt, \nonumber \\
(+)_{N} &\rightarrow& (0)_{N} \mathrm{~with ~probability~} \beta dt, \label{updateright} \\
(+)_{N} &\rightarrow& (-)_{N} \mathrm{~with ~probability~} \gamma dt. \nonumber 
\eea

From the dynamical rules in Eqs. (\ref{updatebulk}-\ref{updateright}), it is clear that the dynamics is symmetric with respect to charge conjugation ($+ \Leftrightarrow -$) combined with space inversion (left $\Leftrightarrow$ right). On setting the boundary exchange rate $\gamma$ to zero, we recover the bridge model studied in \cite{mukamel-ssb1, mukamel-ssb2} which exhibits SSB, as discussed above. In these studies, detailed analysis of the case $q=1$ was carried out. It was also argued that, despite some qualitative changes in the phase diagram as the bulk hopping rate $q$ is varied, the phenomenon of SSB persists for $q \ne 1$. This is consistent with the notion that, in the bridge model, SSB is caused by boundary effects. We thus restrict our investigation of robustness of SSB in this paper to the case $q=1$. Note that in our model, the dynamical moves are the most general ones consistent with the symmetry of the model and with the total asymmetry in the direction of motion of the particles.

In our model, if the exchange rate $\gamma$ equals the particle injection rate $\alpha$, a previous study has demonstrated that no symmetry breaking takes place \cite{schutz-pde}. Moreover, as explained below, if particle extraction occurs \textit{only} by boundary exchange, i.e., if the particle extraction rate $\beta$ equals zero, SSB disappears. Thus, the question arises as to how robust SSB is with respect to the boundary exchange process for general values of $\gamma$. 

In this work, we obtain the complete phase diagram of the model in the space of the three rates $\alpha, \beta, \gamma$, and specify regions where SSB occurs. We base our analysis on a mean-field approximation and an exact analysis in certain parameter regimes. 

We find that for nonzero exchange rate $\gamma$ and for $\beta \ne 0$, the SSB in the bridge model is sustained so long as these rates are not too large. This statement is quantified later in the paper. Thus, SSB in the bridge model is indeed quite robust with respect to additional dynamical steps. Similar to the original model, we find that the average flipping time between the two long-lived states in the symmetry-broken phases grows exponentially with the system size. Our results are corroborated by extensive Monte Carlo simulations of the model.

The paper is organized as follows. In the following section, we discuss how our model is related to the single-species totally asymmetric simple exclusion process (TASEP), and briefly summarize the known phase diagram for the TASEP with open boundary conditions, for use in later parts of the paper. In Section \ref{exact results}, we present exact results for the stationary density profiles and currents for two cases, namely, (i) when the particle extraction rate $\beta$ is zero, and (ii) when the boundary exchange rate $\gamma$ equals the particle injection rate $\alpha$ for non-zero extraction rate $\beta$. In both these cases, the system exhibits symmetric phases only. 

In the absence of an exact solution for the stationary state measure for general values of the system parameters, in Section \ref{mft}, we make progress by applying a mean-field approximation to our model. In particular, we investigate the possibility of various symmetric and symmetry-broken phases in the stationary state. Combined with the results from the previous section, we obtain the complete phase diagram of the model, in which SSB occurs in a certain parameter regime. In Section \ref{MC}, we report extensive Monte Carlo simulations for the particle density profiles in the stationary state, both for the symmetric and the symmetry-broken phases. We also provide numerical evidence for the exponential growth of the average flipping time with system size in the symmetry-broken phases. These results support the mean-field prediction of SSB in our model. In Section \ref{mechanism-ssb}, we briefly describe, with the help of a toy model, the physical mechanism through which symmetry breaking occurs in our model for small values of the particle extraction rate. The paper ends with conclusions in Section \ref{conclusions}.          
\section{Relation of our model to TASEP}
\label{tasep-phasediagram}
Our model generalizes the single-species totally asymmetric simple exclusion process (TASEP) to two species of particles. The TASEP is a paradigmatic model to study nonequilibrium driven systems \cite{asep-rev}. On a periodic one-dimensional lattice, the model involves single species of particles, say, positive, moving stochastically round the lattice by exchanging with nearest-neighbor holes. Defining our model on a periodic lattice with just the bulk dynamics (Eq. (\ref{updatebulk})) and no boundary dynamics, it is clear then that in our model, a positive particle, in its motion round the lattice, will not distinguish between a negative particle and a hole; also, a negative particle will not distinguish between a positive particle and a hole. Thus, for our model on a periodic lattice, the dynamics of the positive and the negative particles becomes that of two separate TASEP's. Indeed, because of these rules, the two particle species (positive and negative) of the two TASEP's behave microscopically entirely independently even though they share the same lattice. The difference between this single-lane model and two non-interacting TASEP's moving on two separate lattices appears only on a coarse-grained level of description through the constraint that the total particle density of positive and negative particles cannot exceed one. Note, however, that the dynamical moves of holes in our model on a periodic lattice do not become that of a TASEP particle. In fact, interpreting the positive particles as regular or first class particles and the negative particles as vacancies, the holes act as the so-called second class particles \cite{second-class}. These second class particles behave like the first class particles in exchanges with vacancies, while they act like the vacancies in exchanges with the first class particles.

In our model on an open lattice with both bulk and boundary dynamics, the two TASEP's of the positive and the negative particles are not entirely independent since the two species interact microscopically at the boundary sites. It turns out that much of the behavior of the model can be deduced from the properties of the single-species TASEP with open boundaries. This process has been well studied in the past, and its phase diagram is exactly known \cite{asep-mft, asep-exact1, asep-exact2}. For use in the later parts of the paper, we briefly summarize below the phase digram of the TASEP in the limit of an infinite system.

Let $\alpha_{s}$ denote the injection rate of particles at the left end of the lattice, while $\beta_{s}$ stands for the extraction rate at the right end of the lattice (Here, the subscript $s$ refers to single species.). The particle exchange rate in the bulk is set to $1$. In the thermodynamic limit, the phase diagram comprises three phases.

\noindent
(i) A maximal current or power law phase for $\alpha_{s} \ge 1/2, \beta_{s} \ge 1/2$. In this phase, the particle density approaches from the boundaries to the bulk value of $1/2$ as a power law, and the current is maximal ($j_{s}=1/4$).

\noindent
(ii) A low-density phase for $\alpha_{s} < \beta_{s}, \alpha_{s} < 1/2$. Here, the particle current is $j_{s}=\alpha_{s}(1-\alpha_{s})$; the bulk density equals $\alpha_{s} (< 1/2)$, and is approached exponentially from the right boundary. 

\noindent
(iii) A high-density phase for $\beta_{s} < \alpha_{s}, \beta_{s} < 1/2$. Here, the particle current is $j_{s}=\beta_{s}(1-\beta_{s})$; the bulk density equals $1-\beta_{s} (>1/2)$, and is approached exponentially from the left boundary.

The transition from the low-density and the high-density phase into the power law phase is continuous, while the transition from the high-density to the low-density phase is discontinuous, with phase coexistence on the line $\alpha_{s}=\beta_{s} < 1/2$ for which the density profile is linear. The linear density profile is a result of superposition of configurations in which a left-hand region of density $\alpha_{s}$ coexists with a right-hand region of density $1-\alpha_{s}$ with a shock between them \cite{asep-mft, asep-exact2, asep-shock}. 
 
\section{The phase diagram: Exact results on specific planes}
\label{exact results}
In this section, we discuss two cases for which exact results for the stationary state current and density profile of particles can be obtained for our model. This is possible because, in these two cases, the dynamics of the system effectively becomes that of either the single-species TASEP or two decoupled TASEP's. The two cases are (i) the $\alpha-\gamma$ plane (i.e., $\beta=0$), and (ii) the plane $\alpha=\gamma$, with $\beta$ non-zero. In both these cases, the system exhibits only symmetric phases in which the magnitude of the average positive current (denoted by $j^{+}$) equals that of the average negative current (denoted by $j^{-}$).
\subsection{The $\alpha-\gamma$ plane with $\beta=0$}
\label{exact-plane1}
\begin{figure}[h!]
\begin{center}
\includegraphics[scale=0.5]{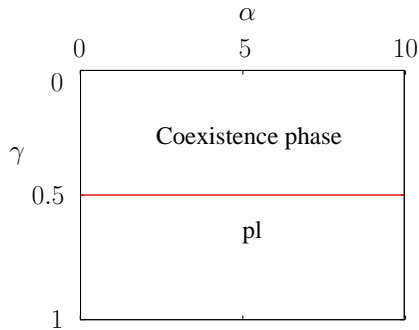}
\caption{The phase diagram on the $\alpha-\gamma$ plane for $\beta=0$, based on the exact analysis in Section \ref{exact-plane1}.}   
\label{phdiag-exact1}
\end{center}
\end{figure}

If the extraction rate $\beta$ is $0$, from the dynamical rules in Eqs. (\ref{updatebulk}-\ref{updateright}), it is evident that holes cannot be injected into the system, and, thus, the system will have just the positive and the negative particles in the stationary state. If one now interprets the negative particles as ``holes'', the dynamics becomes that of a single-species TASEP with the positive particles being injected and extracted with the same rate $\gamma$; the system will thus exhibit the following symmetric phases. (i) For $\gamma \ge 1/2$: Maximal current phase for the positive and the negative particles, with $j^{+}=j^{-}=1/4$. The density for both the particles equals $1/2$ in the bulk, and is approached from the boundaries as a power law. In the spirit of \cite{mukamel-ssb1}, we refer to this phase as the power-law (pl) phase. (ii) For $\gamma < 1/2$: Here, $j^{+}=j^{-}=\gamma(1-\gamma)$, and the density profile is linear for both the species. This phase corresponds to the coexistence line $\alpha_{s}=\beta_{s} < 1/2$ of the single-species TASEP phase diagram. We refer to this phase as the coexistence phase. The schematic phase diagram for the $\alpha-\gamma$ plane with $\beta=0$ is shown in Fig. \ref{phdiag-exact1}. 

\subsection{The plane $\alpha=\gamma$ with $\beta$ non-zero}
\label{exact-plane2}
\begin{figure}[h!]
\begin{center}
\includegraphics[scale=0.5]{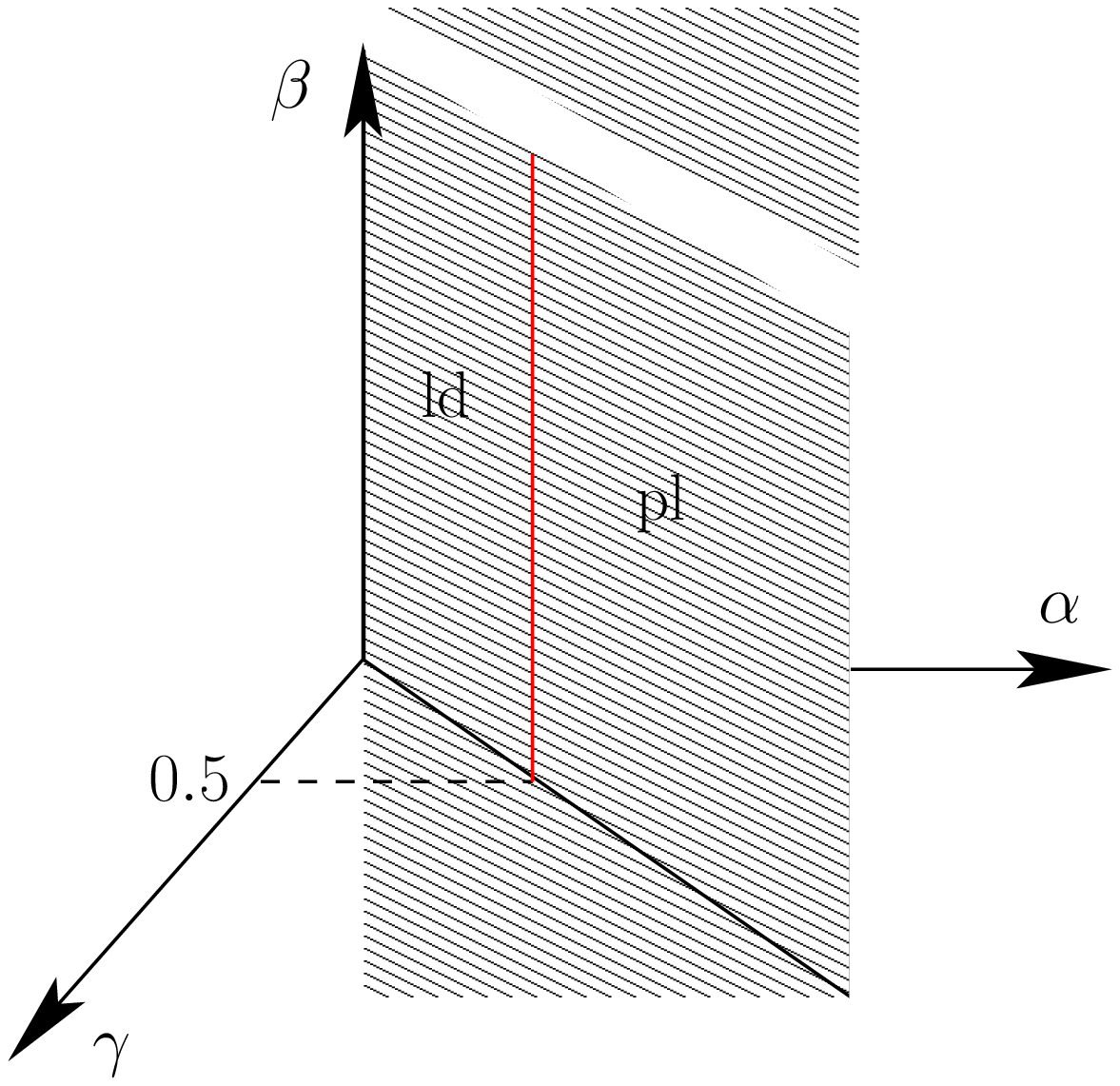}
\caption{The phase diagram on the $\alpha=\gamma$ plane for $\beta \ne 0$, based on the exact analysis in Section \ref{exact-plane2}.}   
\label{phdiag-exact2}
\end{center}
\end{figure}

In this case, the dynamical rules in Eqs. (\ref{updatebulk}-\ref{updateright}) imply that both in the bulk and at the left (respectively, right) boundary, a positive (respectively, negative) particle does not distinguish between a negative (respectively, positive) particle and a hole. As a result, the two TASEP's of the positive and the negative particles are decoupled. The injection rate for both the particles is $\alpha$, while the extraction rate for both is $\alpha+\beta$. The following symmetric phases are possible. (i) For $\alpha \ge 1/2$: Power-law (pl) phase, with $j^{+}=j^{-}=1/4$. The bulk density equals $1/2$ for both the particles and is approached from the two boundaries as a power law. (ii) For $\alpha < 1/2$: Low-density in the bulk for both the positive and the negative particles, with $j^{+}=j^{-}=\alpha(1-\alpha)$. The density of the positive (respectively, negative) particles decays exponentially to the bulk value $\alpha$ as one moves away from the right (respectively, left) boundary. Following \cite{mukamel-ssb1}, we call this the low-density (ld) phase. The transition from the ld to the pl phase is continuous. The schematic phase diagram for the plane $\alpha=\gamma$ with non-zero $\beta$ is shown in Fig. \ref{phdiag-exact2}. 

\section{The Full Phase Diagram: A Mean-field Study}
\label{mft}
In this section, we study the full phase diagram of our model to identify the possible symmetric and symmetry-broken phases. In the absence of an exact solution for the stationary state for general values of the rates $\alpha, \beta, \gamma$, we pursue our study by employing a mean-field approximation. In this approximation, pair and higher-order correlation functions for particle occupations are approximated by products of average occupation numbers.

Let us define two occupation numbers, $\tau_{i}$ and $\theta_{i}$, for each site $i$, where $\tau_{i} ~(\mathrm{respectively}, \theta_{i}) \equiv 1$ if site $i$ has a positive (respectively, negative) particle and is zero otherwise. The hard-core constraint implies only one of $\tau_{i}$ and $\theta_{i}$ to be non-zero at a time.  The occupation number of holes at site $i$ is given by $\sigma_{i} = 1 - \tau_{i} - \theta_{i}$. In this notation, the magnitudes of the positive and the negative currents in the bulk ($ 1 \le i \le N-1 $), derived from Eq. (\ref{updatebulk}) (with $q=1$), are
\bea
j^{+}_{i,i+1}&=&\la \tau_{i}\sigma_{i+1} \ra + \la \tau_{i}\theta_{i+1} \ra=\la \tau_{i}(1-\tau_{i+1}) \ra, \nonumber \\
\label{exactj-b} \\
j^{-}_{i+1,i}&=&\la \theta_{i+1}\sigma_{i} \ra + \la \theta_{i+1}\tau_{i} \ra=\la \theta_{i+1}(1-\theta_{i}) \ra. \nonumber 
\eea
At the two boundaries, the currents (magnitudes only) are given from Eq. (\ref{updateleft}) and Eq. (\ref{updateright}) by
\bea
j^{+}_{0,1}&=&\alpha \la \sigma_{1} \ra + \gamma \la \theta_{1} \ra = \alpha(1-\la \tau_{1} \ra)-(\alpha-\gamma)\la \theta_{1} \ra\nonumber, \\ 
j^{-}_{1,0}&=&(\beta + \gamma) \la \theta_{1} \ra, \nonumber \\
\label{exactj-lr} \\
j^{+}_{N,N+1}&=&(\beta + \gamma) \la \tau_{N} \ra, \nonumber \\ 
j^{-}_{N+1,N}&=&\alpha \la \sigma_{N} \ra + \gamma \la \tau_{N} \ra = \alpha(1-\la \theta_{N} \ra)-(\alpha-\gamma)\la \tau_{N} \ra. \nonumber
\eea

In the above, we have considered only the magnitudes of the current with the understanding that the positive current flows from left to right, while the negative current flows from right to left.  
Let $p_{i}$ (respectively, $m_{i}$) denote the average density of the positive (respectively, negative) particles at site $i$.
\bea
p_{i}&=&\la \tau_{i} \ra, \nonumber \\
\label{dendef-mft} \\
m_{i}&=&\la \theta_{i} \ra. \nonumber  
\eea
These densities evolve in time according to
\bea
\frac{dp_{i}}{dt}&=&j^{+}_{i-1,i}-j^{+}_{i,i+1}, \nonumber \\
\label{denev-mft} \\
\frac{dm_{i}}{dt}&=&j^{-}_{i+1,i}-j^{-}_{i,i-1}. \nonumber  
\eea

We proceed by employing the mean-field approximation, in which the currents on the right hand side of the above equation are replaced by their mean-field values which are obtained from Eq. (\ref{exactj-b}) and Eq. (\ref{exactj-lr}) as
\bea
j^{+}_{i,i+1}&=&p_{i}(1-p_{i+1}), \nonumber \\
\label{jb-mft} \\
j^{-}_{i+1,i}&=&m_{i+1}(1-m_{i}), \nonumber 
\eea
for the bulk $(1 \le i \le N-1)$, and
\bea
j^{+}_{0,1}&=&\alpha(1-p_{1})-(\alpha-\gamma)m_{1}, \nonumber \\
j^{-}_{1,0}&=&(\beta+\gamma)m_{1}, \nonumber \\
\label{j-lr-mft} \\
j^{+}_{N, N+1}&=&(\beta+\gamma)p_{N}, \nonumber \\
j^{-}_{N+1,N}&=&\alpha(1-m_{N})-(\alpha-\gamma)p_{N}, \nonumber 
\eea
for the boundaries.

In the stationary state, the currents of the positive and the negative particles will be constant throughout the system, i.e., $j^{+}_{i,i+1}=j^{+}$ and $j^{-}_{i+1,i}=j^{-}$ for all $i$. As a result, for $1 \le i \le N-1$, one has
\bea
j^{+}&=&p_{i}(1-p_{i+1})\nonumber, \\
\label{j-steady-b} \\
j^{-}&=&m_{i+1}(1-m_{i})\nonumber,
\eea
whereas, at the boundaries, one has
\bea
j^{+}&=&\alpha(1-p_{1})-(\alpha-\gamma)m_{1}=(\beta+\gamma)p_{N}, \nonumber \\
\label{j-steady-lr} \\
j^{-}&=&(\beta+\gamma)m_{1}=\alpha(1-m_{N})-(\alpha-\gamma)p_{N}. \nonumber
\eea
Following \cite{mukamel-ssb1}, we define for the positive species, the effective injection rate $\alpha^{+}$ at the left boundary and the effective extraction rate $\beta^{+}$ at the right boundary in the following way.
\bea
\alpha^{+}&\equiv&\frac{j^{+}}{(1-p_{1})}.\\
\beta^{+}&\equiv&\frac{j^{+}}{p_{N}}.
\eea
From Eq. (\ref{j-steady-lr}), it follows that
\bea
\alpha^{+}&=&\frac{j^{+}}{\frac{j^{+}}{\alpha}+\left(\frac{\alpha-\gamma}{\beta+\gamma}\right)\frac{j^{-}}{\alpha}}, \nonumber \\
\label{effrate+} \\
\beta^{+}&=&\beta+\gamma. \nonumber
\eea
Similarly, one defines for the negative species, the effective injection rate $\alpha^{-}$ at the right boundary and the effective extraction rate $\beta^{-}$ at the left boundary in the following way.
\bea
\alpha^{-}&\equiv&\frac{j^{-}}{(1-m_{N})},\\
\beta^{-}&\equiv&\frac{j^{-}}{m_{1}}.
\eea
On using Eq. (\ref{j-steady-lr}), we get
\bea
\alpha^{-}&=&\frac{j^{-}}{\frac{j^{-}}{\alpha}+\left(\frac{\alpha-\gamma}{\beta+\gamma}\right)\frac{j^{+}}{\alpha}}, \nonumber \\
\label{effrate-} \\
\beta^{-}&=&\beta+\gamma. \nonumber
\eea
Thus, we have two TASEP's of the positive and the negative particles. The positive particles enter at the left boundary with rate $\alpha^{+}$, hop through the bulk with rate $1$, and exit at the right boundary with rate $\beta^{+}$. On the other hand, the negative particles enter at the right boundary with rate $\alpha^{-}$, hop through the bulk with rate $1$, and exit at the left boundary with rate $\beta^{-}$. The two TASEP's are coupled at the boundaries through the effective injection rates $\alpha^{\pm}$, which, for one species, depend on the density of the other. Using the phase diagram of the single-species TASEP, we discuss below the possible phases in our model, first the symmetric ones, followed by those with spontaneous symmetry breaking. In the following, we denote the bulk densities away from the boundaries by $p$ and $m$ for the positive and the negative particles, respectively.
\subsection{Symmetric Phases}
The two possible symmetric phases are the power-law (pl) phase and the low-density (ld) phase. One cannot have a high-density symmetric phase, since, in that case, the combined bulk density of particles would exceed one. In these symmetric phases, one has $j^{+}=j^{-}$ ($=j^{s}$, say) and therefore, from  Eq. (\ref{effrate+}) and Eq. (\ref{effrate-}), one has
\be
\alpha^{+}=\alpha^{-}=\frac{\alpha(\beta+\gamma)}{\alpha+\beta}=\alpha^{s}, \mathrm{~say},
\ee
while one already has $\beta^{+}=\beta^{-}$ in the definition of the model. 
\begin{itemize}
\item{\textbf{pl phase:} 

Here, $j^{s}=1/4$, and far away from the boundaries, the densities are $p=m=1/2$. The conditions for the occurrence of this phase are
\be
\alpha^{s} \ge 1/2, \beta+\gamma \ge 1/2.
\ee
Now, $\beta+\gamma$ is always greater than or equal to $\alpha^{s}$, so that there is a single condition defining this phase, namely,
\be
\alpha^{s}\ge \frac{1}{2} \mathrm{~(pl ~phase)}.
\label{cond-pl}
\ee
}
\item{\textbf{ld phase:}

In this phase, $j^{s}=\alpha^{s}(1-\alpha^{s})$, and the bulk densities are given by $p=m=\alpha^{s}$. This phase exists provided the following conditions are satisfied.
\be
\alpha^{s} < \beta+\gamma, \alpha^{s} < 1/2.
\ee
The first condition is always satisfied, so that the single condition defining this phase is
\be
\alpha^{s} < 1/2 \mathrm{~(ld ~phase)}.
\label{cond-ld}
\ee
}
\end{itemize}
From Eq. (\ref{cond-pl}) and Eq. (\ref{cond-ld}), it follows that the surface in the $(\alpha, \beta, \gamma)$ space that marks the transition between the two symmetric phases is given by 
\be
\alpha^{s}=\frac{\alpha(\beta+\gamma)}{\alpha+\beta} = \frac{1}{2}.
\ee
Solving for $\beta$, one gets 
\be
\beta=\frac{\alpha(1-2\gamma)}{2\alpha-1}.
\ee

The phase transition from the ld to the pl phase is continuous, since, across the transition point, the current $j^{s}$ and its first derivative with respect to density are continuous, while the second derivative is not.

\subsection{Symmetry-broken Phases}
A symmetry-broken phase necessarily means unequal densities for the positive and the negative particles. Then, the different possible symmetry-broken phases are given by the six combinations of the single-species TASEP phases, summarized in Section \ref{tasep-phasediagram}, for the positive and the negative particles. Of these, the high-density/high-density phase and the high-density/power-law phase cannot occur, since for these combinations, the combined bulk densities of the positive and the negative particles would be greater than $1$. It can be shown that the power-law/low-density phase does not exist within the mean-field approximation; the proof is outlined in \ref{app1}. Thus, the possible symmetry-broken phases are the high-density/low-density (hd/ld) phase and the low-density/low-density (ld/ld) phase. Below we investigate the possibility of occurrence of these phases within the mean-field approximation. We assume, without loss of generality, that the positive particles are in the majority in these phases. 
\begin{itemize}
\item{\textbf{hd/ld phase:}
Here, we have $j^{+}=(\beta+\gamma)(1-\beta-\gamma)$ and $j^{-}=\alpha^{-}(1-\alpha^{-})$. The bulk densities are given by $p=1-\beta-\gamma, m=\alpha^{-}$.
The conditions for the existence of this phase are
\bea
\alpha^{+} > \beta + \gamma,&& \beta + \gamma < 1/2, \nonumber \\
\label{cond-hdld} \\
\alpha^{-} < \beta + \gamma,&& \alpha^{-} < 1/2. \nonumber 
\eea
}
\item{\textbf{ld/ld phase:}

Here, both the positive and the negative particles have unequal bulk densities of values smaller than $1/2$. In this phase, we have $j^{+}=\alpha^{+}(1-\alpha^{+})$ and $j^{-}=\alpha^{-}(1-\alpha^{-})$. The bulk densities are given by $p=\alpha^{+}, m=\alpha^{-}$. With the positive particles in the majority, one has $\alpha^{+}>\alpha^{-}$, and consequently, $j^{+} > j^{-}$. The conditions for the existence of this phase are
\bea
\alpha^{+} < \beta + \gamma,&& \alpha^{+} < 1/2, \nonumber \\
\label{cond-ldld} \\
\alpha^{-} < \beta + \gamma,&& \alpha^{-} < 1/2. \nonumber 
\eea
}
\end{itemize}

Comparing the conditions in Eqs. (\ref{cond-hdld}-\ref{cond-ldld}), it follows that the transition surface between the hd/ld phase and the ld/ld phase is given by
\be
\alpha^{+}=\beta+\gamma.
\ee
To plot the above surface, one has to express $\alpha^{+}$ in terms of $\alpha, \beta$, and $\gamma$. In the hd/ld phase, one has, from Eq. (\ref{effrate-}), on substituting $j^{+}=(\beta+\gamma)(1-\beta-\gamma)$ and $j^{-}=\alpha^{-}(1-\alpha^{-})$, 
\be
\alpha^{-}=\frac{\alpha^{-}(1-\alpha^{-})}{\frac{\alpha^{-}(1-\alpha^{-})}{\alpha}+\left(\frac{\alpha-\gamma}{\beta+\gamma}\right)\frac{(\beta+\gamma)(1-\beta-\gamma)}{\alpha}},
\ee  
which gives a quadratic equation in $\alpha^{-}$. Solving this equation, we get 
\be
\alpha^{-}=\frac{1+\alpha}{2}-\frac{1}{2}\sqrt{(1+\alpha)^{2}-4\{\alpha(\beta+\gamma)+\gamma(1-\beta-\gamma)\}},
\label{alpha-}
\ee
where the root with the negative sign is taken to ensure that $\alpha^{-} < 1/2$. The above equation gives the value of $\alpha^{-}$ throughout the hd/ld and the ld/ld phase.

In order to get an expression for $\alpha^{+}$ for the hd/ld phase, we substitute in Eq. (\ref{effrate+}) the currents $j^{+}=(\beta+\gamma)(1-\beta-\gamma)$ and $j^{-}=\alpha^{-}(1-\alpha^{-})$ for this phase; we get 
\be
\alpha^{+}=\frac{(\beta+\gamma)(1-\beta-\gamma)}{\frac{(\beta+\gamma)(1-\beta-\gamma)}{\alpha}+\left(\frac{\alpha-\gamma}{\beta+\gamma}\right)\frac{\alpha^{-}(1-\alpha^{-})}{\alpha}}.
\label{alpha+}
\ee 
For the ld/ld phase, the expression for $\alpha^{+}$ follows from Eq. (\ref{effrate+}) on substituting the currents for this phase, namely, $j^{+}=\alpha^{+}(1-\alpha^{+})$ and $j^{-}=\alpha^{-}(1-\alpha^{-})$; one gets
\be
\alpha^{+}=\frac{\alpha^{+}(1-\alpha^{+})}{\frac{\alpha^{+}(1-\alpha^{+})}{\alpha}+\left(\frac{\alpha-\gamma}{\beta+\gamma}\right)\frac{\alpha^{-}(1-\alpha^{-})}{\alpha}}.
\label{alpha+ldld}
\ee
The above equation leads to a quadratic equation in $\alpha^{+}$, which is solved and the root with the negative sign, satisfying $\alpha^{+} <1/2$, gives $\alpha^{+}$ in the ld/ld phase.

From Eqs. (\ref{alpha+}) and (\ref{alpha+ldld}), it is clear that the expressions for $\alpha^{+}$ for the hd/ld and the ld/ld phases become identical when the condition $\alpha^{+}=\beta+\gamma$ is satisfied; this condition thus gives the transition surface between the two phases, as already discussed above. In order to get an explicit expression for this surface in terms of $\alpha, \beta, \gamma$, one may proceed as follow. Inserting the expression for $\alpha^{-}$ from Eq. (\ref{alpha-}) into Eq. (\ref{alpha+}), evaluating $\alpha^{+}$ and then equating $\alpha^{+}$ to $\beta+\gamma$, we get a cubic equation in $\alpha$ in terms of $\beta$ and $\gamma$. We solve this equation, and choose the particular root which gives $\beta$ as a function of $\alpha$ and $\gamma$ such that the condition, $\beta+\gamma<1/2$, is satisfied; we finally get the equation of the surface separating the hd/ld phase from the ld/ld phase. 

The existence of the ld/ld phase in our model is demonstrated within the mean-field approximation. In the original bridge model, whether this phase exists beyond the mean-field approximation has been a subject of some debate \cite{ahr-ssb,clincy-ssb,zia-ssb}.   

It may be checked from Eqs. (\ref{effrate+}), (\ref{effrate-}), (\ref{alpha-}), (\ref{alpha+}) and (\ref{alpha+ldld}) that in the regions corresponding to the intersection of the hd/ld phase and the ld/ld phase with the $\beta=0$ plane, one has $\alpha^{+}=\alpha^{-}=\beta^{+}=\beta^{-}=\gamma$. Thus, there is no physical significance of the intersection of the transition surface, $\alpha^{+}=\beta+\gamma$, with the $\beta=0$ plane. This is because, on either side of the intersection curve $\alpha^{+}=\gamma$, the injection rate and the extraction rate of both the positive and the negative particles equal $\gamma$. This is consistent with the observations in Section \ref{exact-plane1}, and consequently, one has only symmetric phases on the $\beta=0$ plane.   
\subsection{Transition between the ld and the ld/ld phases}
From Eq. (\ref{effrate+}) and Eq. (\ref{effrate-}), on substituting the current for the ld/ld phase, namely, $j^{+}=\alpha^{+}(1-\alpha^{+})$ and $j^{-}=\alpha^{-}(1-\alpha^{-})$, we get
\bea
\alpha^{+}&=&1-\frac{1}{\alpha}\alpha^{+}(1-\alpha^{+})-\frac{1}{\beta^{*}}\alpha^{-}(1-\alpha^{-}), \nonumber \\
\label{ld-ldld1} \\
\alpha^{-}&=&1-\frac{1}{\beta^{*}}\alpha^{+}(1-\alpha^{+})-\frac{1}{\alpha}\alpha^{-}(1-\alpha^{-}), \nonumber
\eea
where 
\be
\beta^{*}\equiv\frac{\alpha(\beta+\gamma)}{\alpha-\gamma}.
\ee
To solve Eq. (\ref{ld-ldld1}), we follow the procedure in \cite{mukamel-ssb2}, and define
\be
S=\alpha^{+}+\alpha^{-}; D=\alpha^{+}-\alpha^{-}.
\label{ld-ldld2}
\ee
On taking the difference of the two equations in (\ref{ld-ldld1}), we get
\be
D=\left(\frac{\alpha-\beta^{*}}{\alpha\beta^{*}}\right)D(1-S).
\ee
Now, since we are dealing with a symmetry-broken phase, $D \ne 0$. Thus, one has
\be
S=1-\frac{\alpha\beta^{*}}{\alpha-\beta^{*}}.
\label{ld-ldld3}
\ee
Summing the two equations in (\ref{ld-ldld1}), we get
\be
D=\left[(S-2)\left(\frac{2\alpha\beta^{*}}{\alpha+\beta^{*}}-S\right)\right]^{1/2}.
\ee
At the transition to the ld phase, the densities become equal so that $D=0$. Thus, either $S=2$, which is excluded, since this means that the sum of the bulk densities of particles is greater than $1$, or, that
\be
S=\frac{2\alpha\beta^{*}}{\alpha+\beta^{*}}.
\ee
The above equation, combined with Eq. (\ref{ld-ldld3}), gives the following expression for the rates $\alpha^{+}, \alpha^{-}$ on the transition surface between the ld/ld phase and the ld phase.
\be
\alpha^{+}=\alpha^{-}=\frac{\alpha\beta^{*}}{\alpha+\beta^{*}}=\frac{1}{2}\left(1-\frac{\alpha\beta^{*}}{\alpha-\beta^{*}}\right).
\label{ld-ldld-transition}
\ee

Solving the above equation for $\beta$, we get 
\be
\beta=\frac{-3 \alpha^2+5 \alpha \gamma+\left(\sqrt{9 \alpha^{2}-4 \alpha+4}\right) (\alpha-\gamma)-2 \gamma}{2 (1-\alpha)}.
\label{ld-ldld-transition1}
\ee
In the limit $\alpha \rightarrow 1$,  we get
\be
\beta=\frac{1-4\gamma}{3}.
\ee 
The intersection of the transition surface in Eq. (\ref{ld-ldld-transition1}) with the $\beta=0$ plane is along the curve
\be
\alpha=\frac{2(2\gamma^{2}-\gamma)}{3\gamma-1}.
\ee

Note, however, that there is no physical significance of the above intersection curve. This is because one can check that on either side of this intersection curve, the injection rate and the extraction rate of both the positive and the negative particles are equal to $\gamma$. This observation is consistent with that in Section \ref{exact-plane1}, and consequently, one has only symmetric phases on the $\beta=0$ plane. 

\subsection{Summary of the phase diagram}
We give below the equations of the three surfaces separating the various phases within the mean-field theory.
\begin{itemize}
\item{
Surface that separates the ld phase from the pl phase:
\be
\beta=\frac{\alpha(1-2\gamma)}{2\alpha-1}.
\label{ld-pl-sur}
\ee
}
\item{
Surface that separates the ld phase from the ld/ld phase:
\bea
\beta&=&\frac{1-4\gamma}{3} \mathrm{~for~} \alpha=1. \nonumber \\
\label{ld-ldld-sur} \\
      &=&\frac{-3 \alpha^2+5 \alpha \gamma+\left(\sqrt{9 \alpha^{2}-4 \alpha+4}\right) (\alpha-\gamma)-2 \gamma}{2 (1-\alpha)} \nonumber \\
      && \mathrm{for~} \alpha \ne 1. \nonumber 
\eea
}
\item{
Surface that separates the ld/ld phase from the hd/ld phase:
\be
\alpha^{+}=\beta+\gamma,
\label{hdld-ldld-sur}
\ee
where $\alpha^{+}$ may be found from Eqs. (\ref{alpha-}) and (\ref{alpha+}).
}
\end{itemize}

\begin{figure}[h!]
\begin{center}
\includegraphics[scale=0.4]{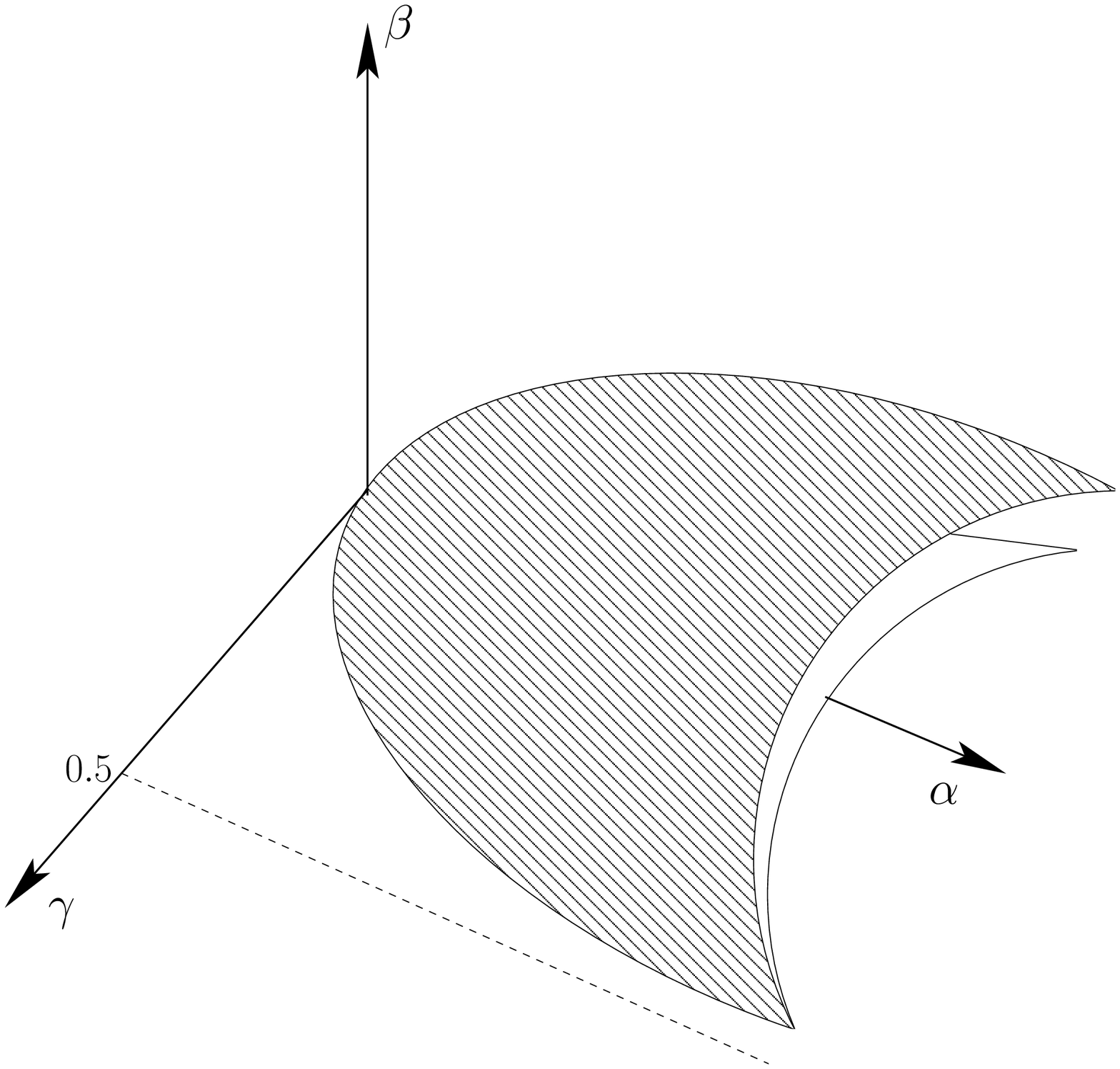}
\caption{Schematic phase diagram in the $\alpha-\beta-\gamma$ space, showing only the regions occupied by the symmetry-broken phases, as predicted by the mean-field approximation. Within the region enclosed by the inner surface, one has the hd/ld phase, while, within the region between the inner surface and the outer shaded one, one has the ld/ld phase. The two surfaces intersect on the $\beta=0$ plane. As explained in the text, the hd/ld phase extends up to but not including the $\beta=0$ plane.}
\label{3dphdiag}
\end{center}
\end{figure}

\begin{figure}[h!]
\begin{center}
\includegraphics[scale=0.5]{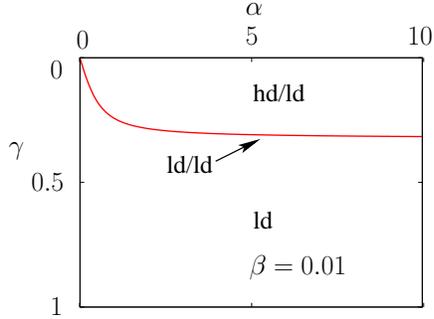}
\caption{The phase diagram on the $\alpha-\gamma$ plane for $\beta=0.01$, based on the mean-field analysis in Section \ref{mft}. The ld/ld phase occupies a very narrow region, and hence, appears as a line on the scale of the figure.}   
\label{phdiag-1}
\end{center}
\end{figure}

\begin{figure}
\begin{center}
\includegraphics[scale=0.5]{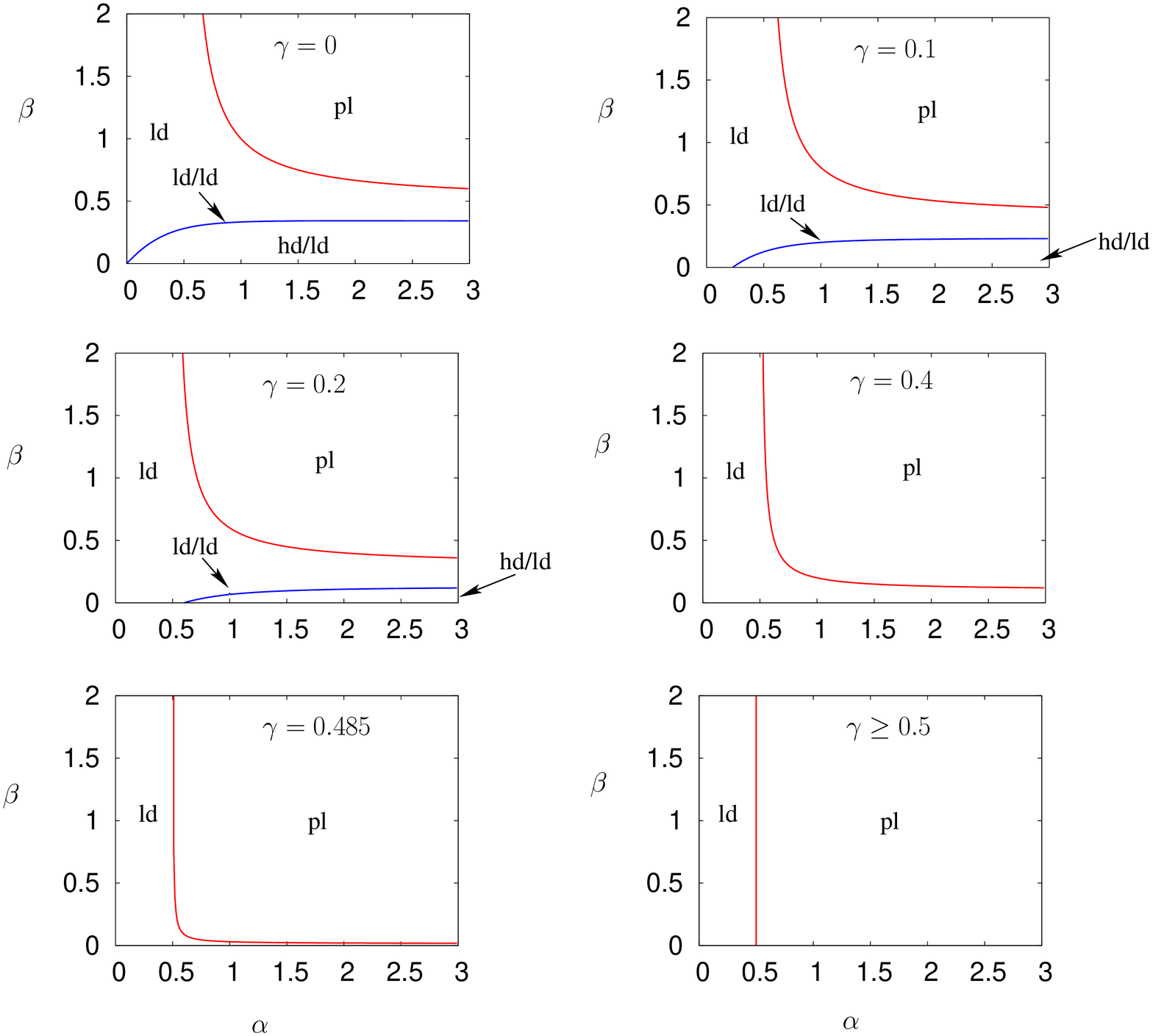}
\caption{The figure shows the mean-field phase diagram on the $\alpha-\beta$ plane for six values of $\gamma$. The phase boundaries are given by Eqs. (\ref{ld-pl-sur}-\ref{hdld-ldld-sur}). The ld/ld phase occupies a very narrow region which appears as a line on the scale of the plots.}
\label{phdiag-2}
\end{center}
\end{figure}

Note that the two transition surfaces, given by Eqs. (\ref{ld-ldld-sur}) and (\ref{hdld-ldld-sur}), coincide on the $\beta = 0$ plane. This can be seen in the schematic phase diagram in the $\alpha-\beta-\gamma$ space, given in Fig. \ref{3dphdiag}. In this phase diagram, we show only the regions occupied by the symmetry-broken phases, as predicted by the mean-field approximation to our model.  In Fig. \ref{phdiag-1}, we show the phase diagram on the $\alpha-\gamma$ plane for $\beta=0.01$, obtained from our mean-field analysis. In Fig. \ref{phdiag-2}, we show the mean-field phase diagram on the $\alpha-\beta$ plane for six values of $\gamma$. 

The transition from the hd/ld phase to the ld/ld phase is discontinuous, while that from the ld/ld phase to the ld phase is continuous. This is because, in the former case, the first derivative of the average current with respect to density is discontinuous across the transition, while, in the latter case, the second derivative of the current with respect to density changes discontinuously across the transition.   

\section{Monte Carlo simulations}
\label{MC}
To check the general features of the phase diagram predicted by the mean-field theory, and in particular, the existence of the symmetry-broken phases, we carried out extensive Monte Carlo (MC) simulations of our model. For a lattice of size $N$ and given values of the parameters $\alpha, \beta$ and $\gamma$, we followed a random sequential update scheme for the configurations, with $N+1$ updates constituting one Monte Carlo Step (MCS). We typically ran the simulation for sufficiently long time ($\sim N^{3}$ MCS) to ensure that the system relaxes to stationarity in this time, after which we started making measurements for the average density profiles and the currents for the two species of particles. 

In the symmetry-broken phases, in order to get the average density profiles, the occupation of each site was averaged over several runs of the simulation in the stationary state. However, one has to be careful that the running time of the simulations does not exceed the flipping time $\tau(N)$ between the two states of the symmetry-broken phases. An estimate of $\tau(N)$ may be made following a procedure explained later in this section. In the symmetric phases, no such restriction on the running time is necessary so that one runs the simulation long enough to reduce fluctuations in the measured density profiles. Following \cite{mukamel-ssb2}, the currents are measured as
\be
j^{\pm}=\frac{N_{\pm}}{(N+1)N_{\mathrm{st}}},
\ee
where $N_{\pm}$ is the total number of positive (negative) particles which have moved in $N_{\mathrm{st}}$ MCS/site. 

The Monte Carlo density profiles for the symmetry-broken phases are shown in Figs. \ref{df-hd/ld phase} and \ref{df-ld/ld phase} corresponding to the hd/ld phase and the ld/ld phase, respectively, while those for the symmetric phases are shown in Figs. \ref{df-ld phase} and \ref{df-pl phase} corresponding to the ld phase and the pl phase, respectively. From the figures, it can be seen that the density profiles are flat in the bulk, with some structures near the boundaries. In the hd/ld phase, the density of the positive particles is higher than $1/2$, while the density of the negative particles is lower than $1/2$; in the ld/ld phase, both densities are smaller than $1/2$ and unequal in magnitude (shown in the blowup of the bulk density profiles). In the symmetric phases, on the other hand, the densities of both particle types are equal, taking a value which is either smaller than $1/2$ (ld phase), or, equal to $1/2$ (pl phase).
  
\bef
\begin{center}
\includegraphics[scale=0.58]{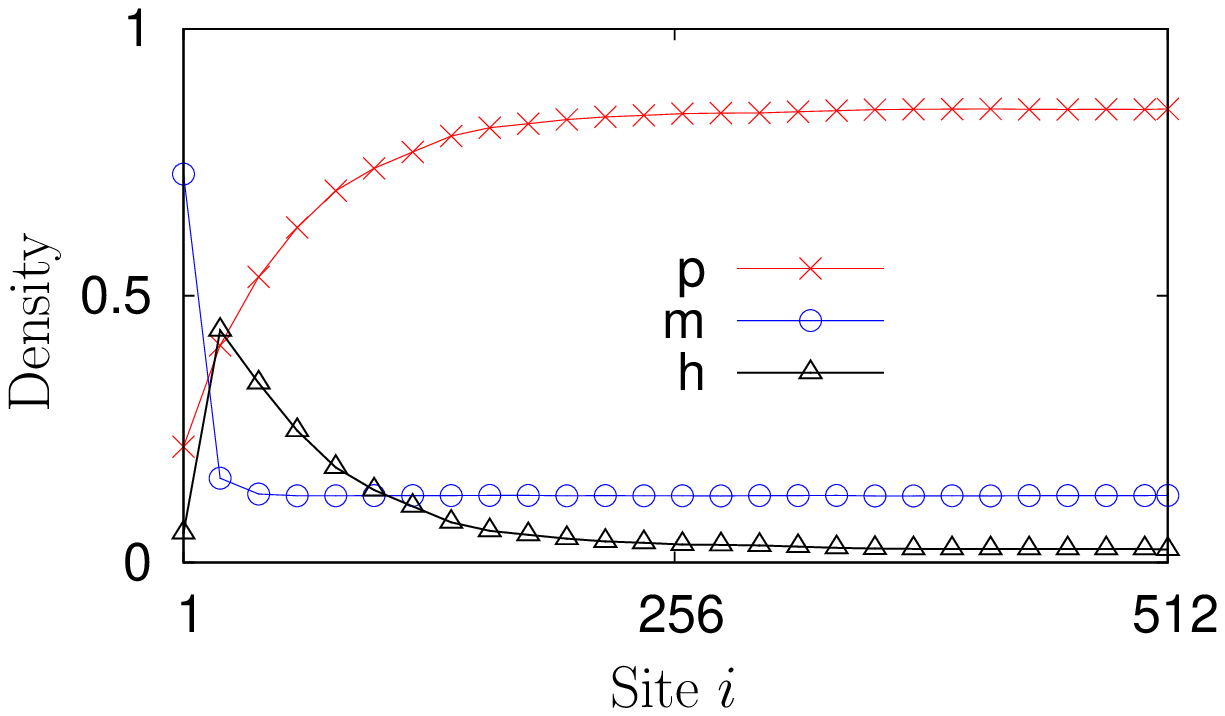}
\caption{Monte Carlo density profiles for the hd/ld phase. Legends: p $\rightarrow$ positive particles, m $\rightarrow$ negative particles, h $\rightarrow$ holes. Here, $\alpha=1.0,\beta=0.05,\gamma=0.1, N=512$.}
\label{df-hd/ld phase}
\end{center}
\eef
\bef
\begin{center}
\includegraphics[scale=0.58]{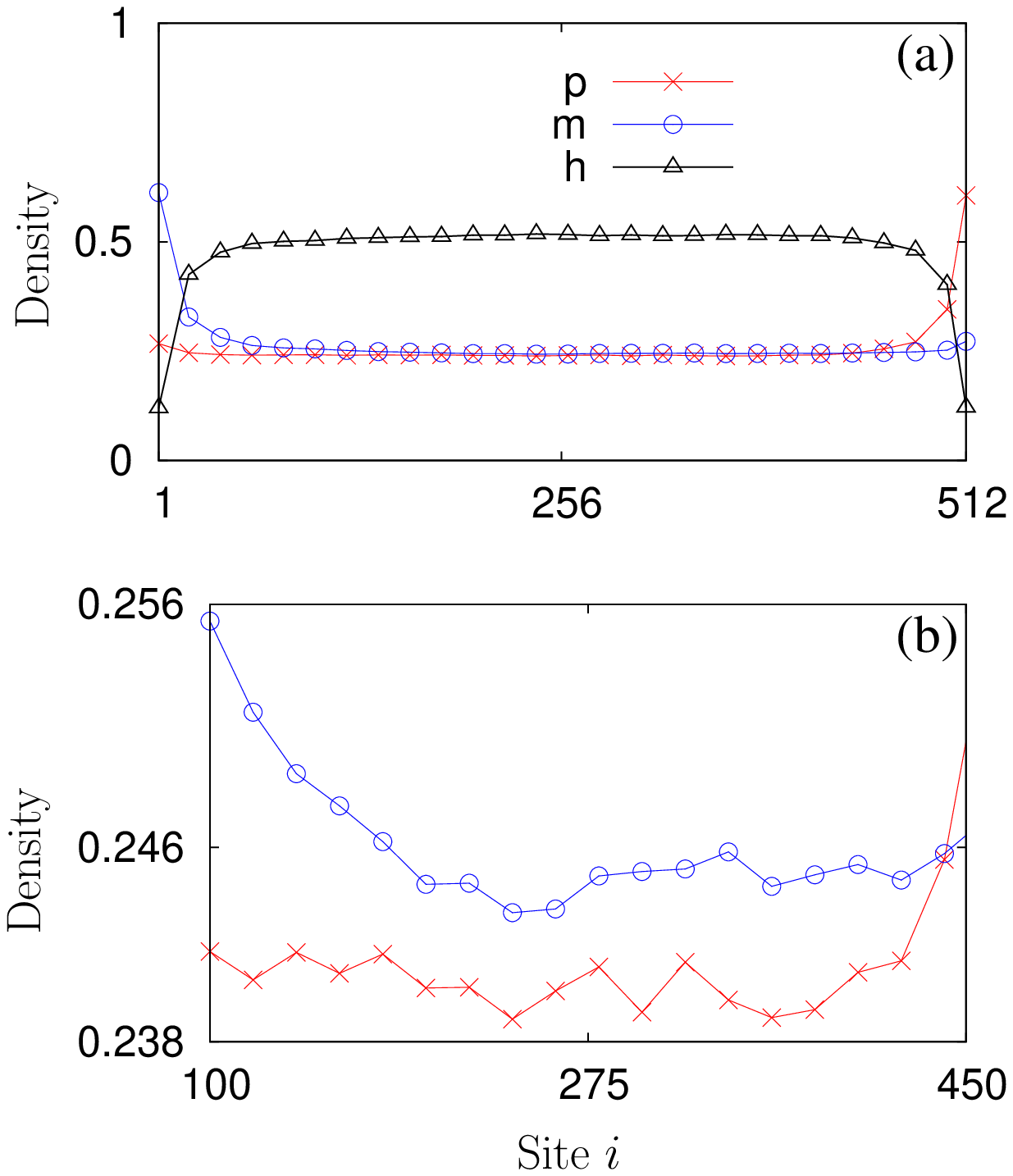}
\caption{(a) Monte Carlo density profiles for the ld/ld phase. Legends: p $\rightarrow$ positive particles, m $\rightarrow$ negative particles, h $\rightarrow$ holes. Here, $\alpha=1.0,\beta=0.1984,\gamma=0.1, N=512$. (b) A blowup of the particle density profiles in the bulk, showing unequal values of the average density for the positive and the negative particles.}
\label{df-ld/ld phase}
\end{center}
\eef
\bef
\begin{center}
\includegraphics[scale=0.58]{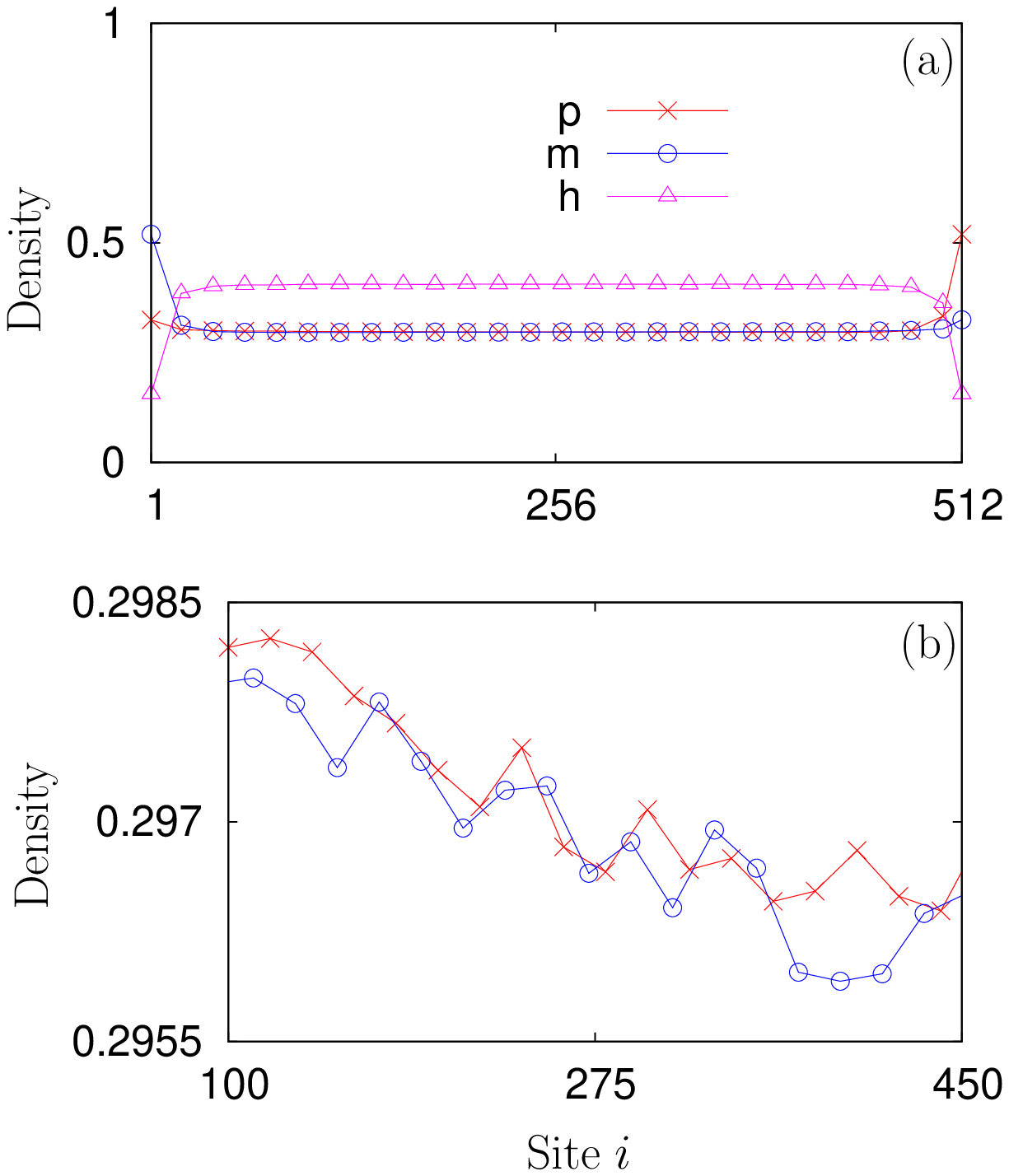}
\caption{(a) Monte Carlo density profiles for the ld phase. Legends: p $\rightarrow$ positive particles; m $\rightarrow$ negative particles, h $\rightarrow$ holes. Here, $\alpha=1.0,\beta=0.3,\gamma=0.1, N=512$. (b) A blowup of the particle density profiles in the bulk. Here, we plot $p_i$ and $m_{N-i}$ to demonstrate that charge conjugation combined with space inversion symmetry is not broken in this phase.}
\label{df-ld phase}
\end{center}
\eef
\bef
\begin{center}
\includegraphics[scale=0.58]{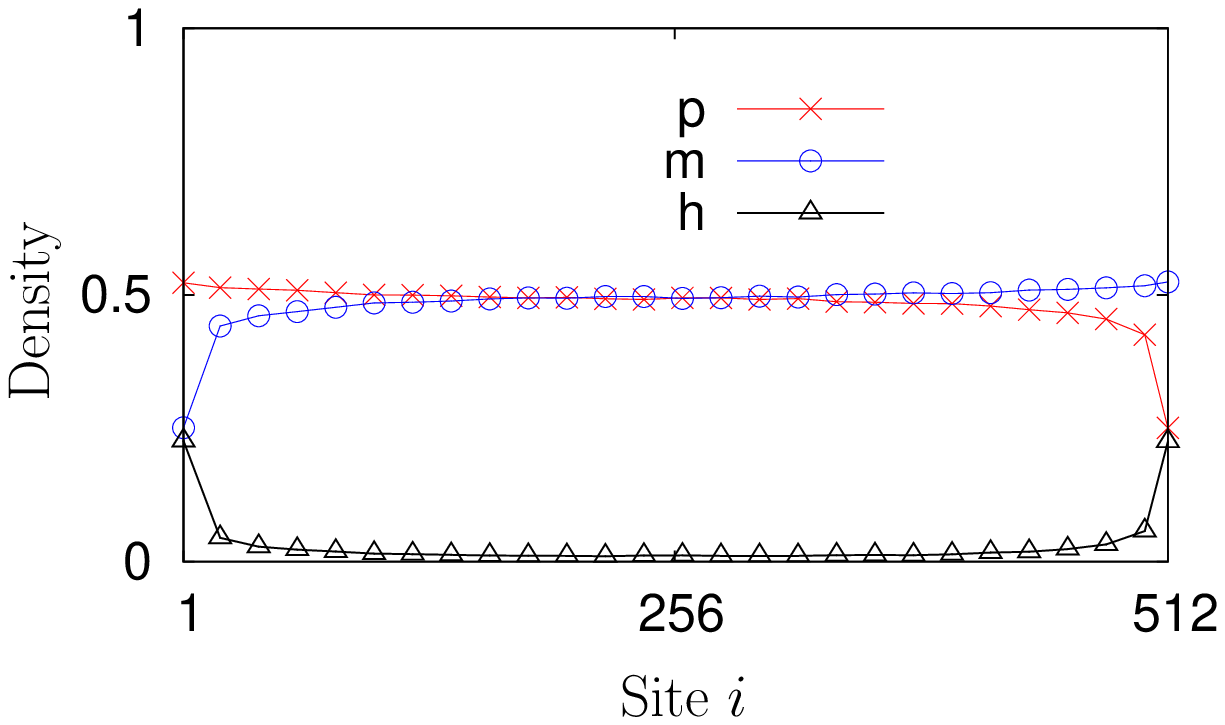}
\caption{Monte Carlo density profiles for the pl phase. Legends: p $\rightarrow$ positive particles, m $\rightarrow$ negative particles, h $\rightarrow$ holes. Here, $\alpha=1.0,\beta=0.9,\gamma=0.1, N=512$.}
\label{df-pl phase}
\end{center}
\eef

In order to illustrate symmetry breaking in our model, we present in Fig. \ref{flips in J} the time evolution of the current difference $j^{+}-j^{-}$ for a typical run of the MC simulation in the hd/ld phase. From the figure, it is evident that, excepting for short time intervals during which flips take place, the system is loaded predominantly with either the positive or the negative particles, alternating between the two as time progresses. We now proceed to show that the average time $\tau(N)$ between successive flips of the current difference grows exponentially with the system size $N$. In order to estimate $\tau(N)$ numerically, we averaged the current difference over many runs, starting from the configuration where all sites are occupied by one species of particles, say, the positive particles only \cite{mukamel-ssb1}. This averaged quantity decays in time because of flips in current difference as a function of time. At large time $t$, in a system of size $N$, the average current difference decays as $\exp[-t/\tau(N)]$, which gives the flipping time $\tau(N)$. 

The results for $\tau(N)$, extracted from MC simulations for various $N$ and given values of $\alpha, \beta, \gamma$, is shown in Fig. \ref{tauN vs. N}. It is readily seen that the plot asymptotically becomes linear, indicating that the time scale $\tau(N)$ grows as an exponential in the system size $N$ for large $N$. This fact, combined with our observation for flat density profiles in the bulk for the symmetry-broken phases, leads us to conclude that in the relevant parameter regimes (see Fig. \ref{3dphdiag}), our model exhibits spontaneous symmetry breaking.        
\bef
\begin{center}
\includegraphics[scale=0.6]{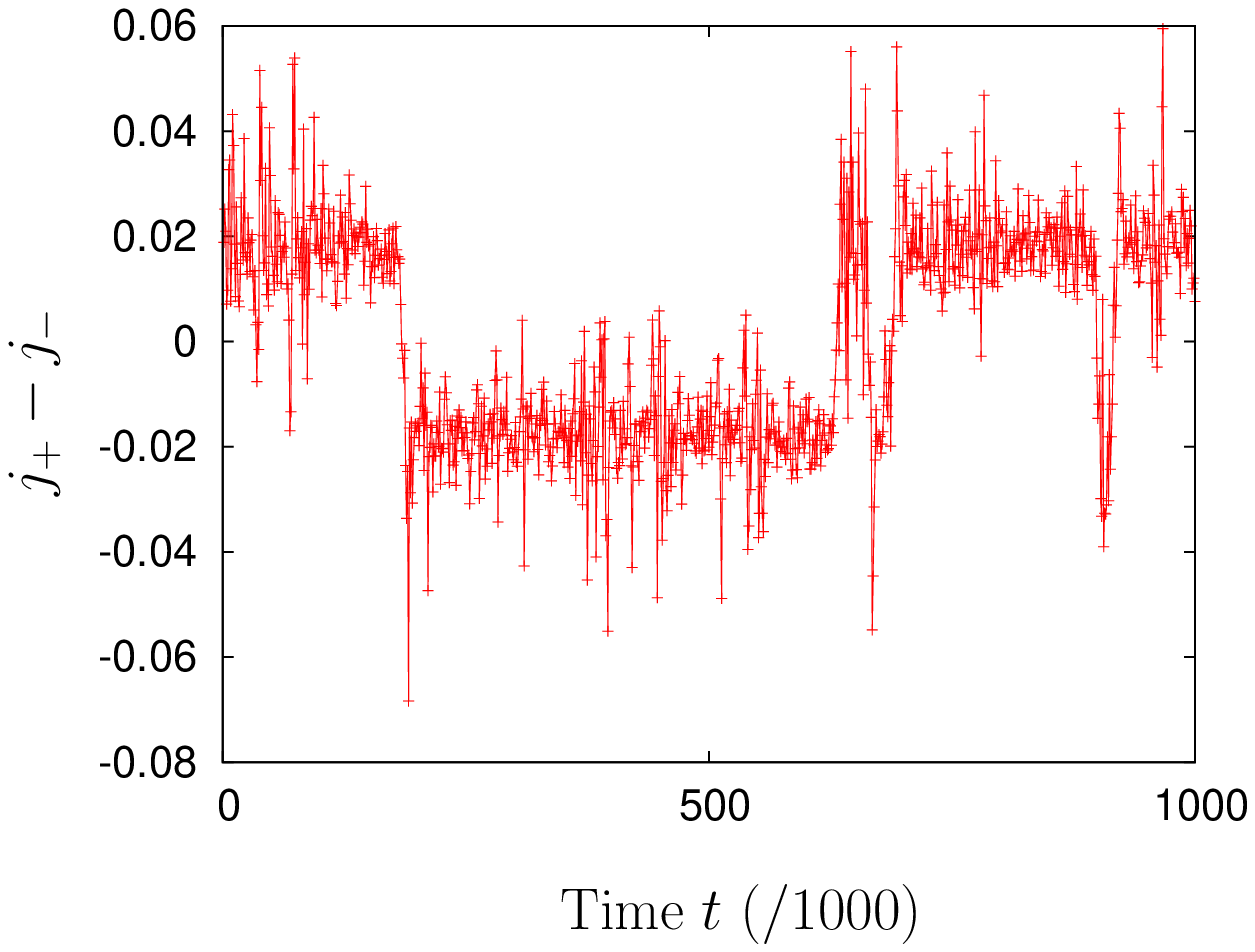}
\caption{Time evolution of the current difference in a typical run of the Monte Carlo simulation in the hd/ld phase. Here, $\alpha=1.0,\beta=0.05,\gamma=0.1, N=256$. Each point represents an average of the current difference over $1000$ Monte Carlo sweeps.} 
\label{flips in J}
\end{center}
\eef
\bef
\begin{center}
\includegraphics[scale=0.6]{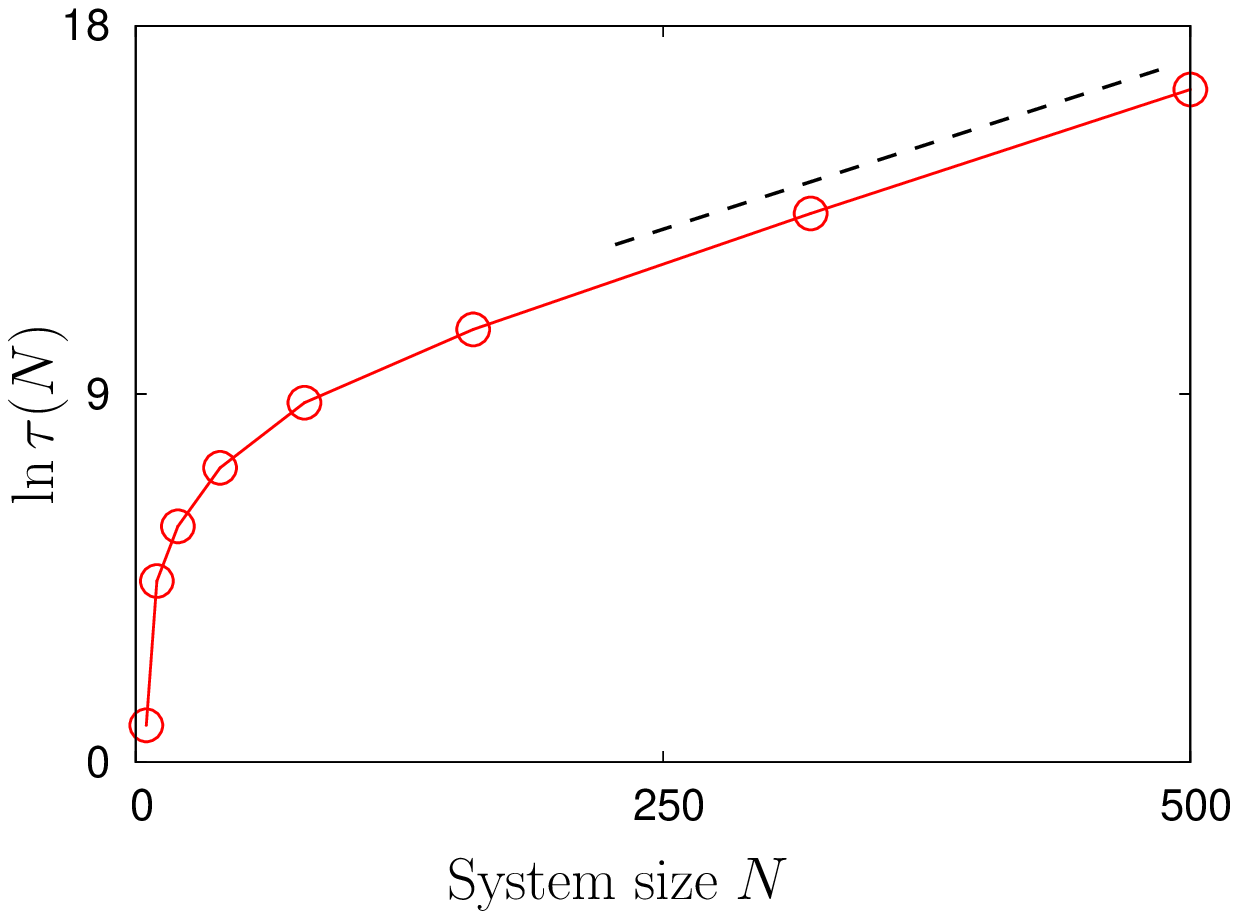}
\caption{Average flipping time $\tau(N)$ as a function of system size $N$. Here, $\alpha=1.0,\beta=0.05,\gamma=0.1$. The points are obtained from Monte Carlo simulations. The dashed line is to indicate that the plot asymptotically becomes linear.}  
\label{tauN vs. N}
\end{center}
\eef

In passing, we note that for fixed and finite $N$, in the two symmetry-broken phases, as time passes, the average density profiles and currents for the particles would appear progressively symmetric due to repeated flips between the two symmetry-related states. Hence, while identifying symmetry-broken phases, it would be more appropriate to look at symmetric combinations of currents or densities, e.g., the sum and the absolute difference of currents \cite{mukamel-ssb2}. Figure \ref{jsum-jdiff} shows $(j^{+}+j^{-})/2$ and $|j^{+}-j^{-}|/2$ as a function of $\beta$ for $\alpha=1.0, \gamma=0.1$. The points are obtained from MC simulations of a system of size $320$, while the continuous lines are the mean-field predictions of Section \ref{mft}. One finds a fairly good agreement for the sum, although not for the difference.
\bef
\begin{center}
\includegraphics[scale=0.6]{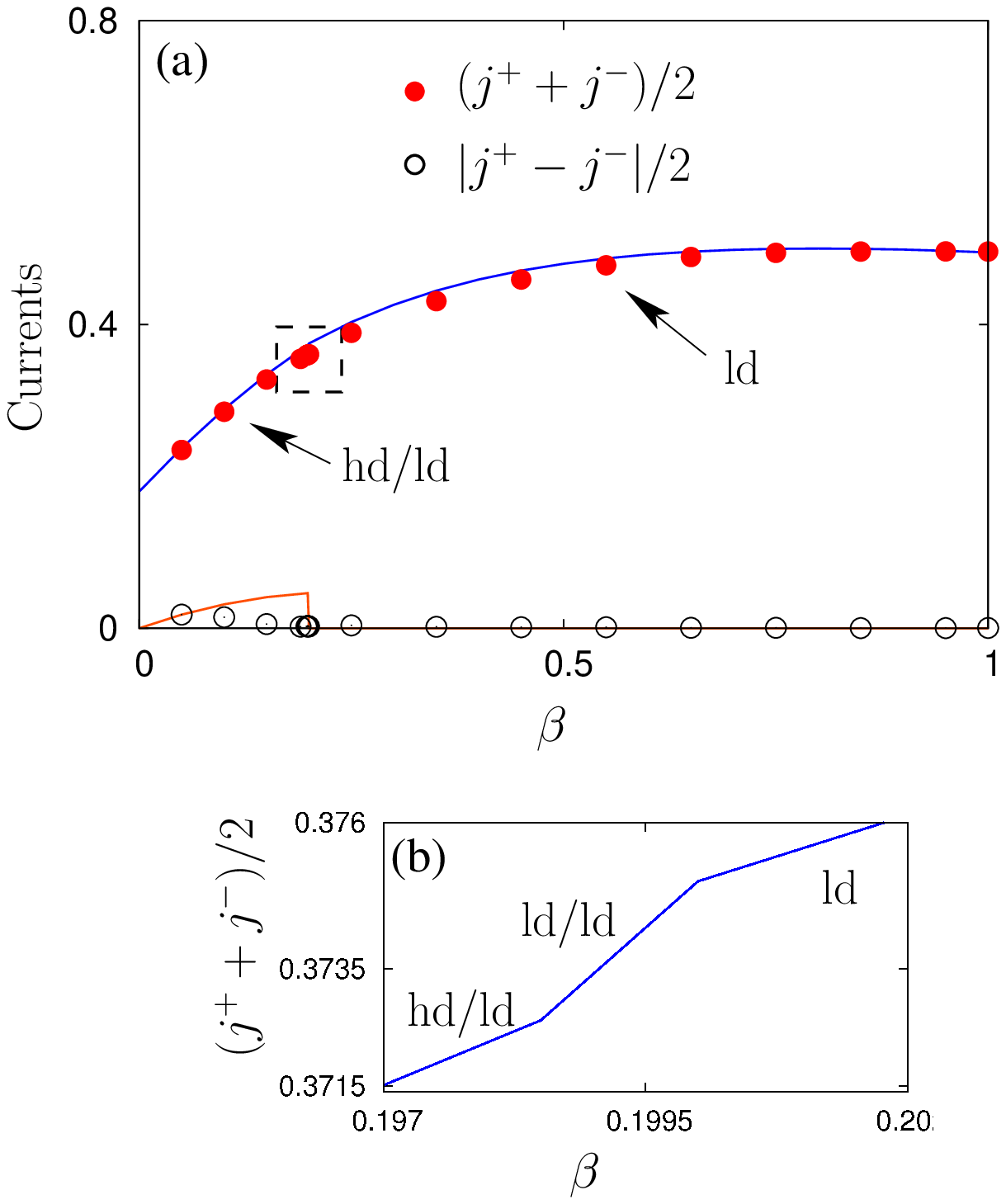}
\caption{(a) Currents $(j^{+}+j^{-})/2$ and $|j^{+}-j^{-}|/2$ as a function of $\beta$ for $\alpha=1.0, \gamma=0.1, N=320$. The continuous lines represent the mean-field results, while points correspond to Monte Carlo simulation results. (b) Blowup of the dotted box in (a), showing the mean-field behavior of $(j^{+}+j^{-})/2$ in the transition region from the hd/ld to the ld/ld phase.}
\label{jsum-jdiff}
\end{center}
\eef
\section{Mechanism for symmetry breaking}
\label{mechanism-ssb}
In this section, we briefly discuss the utility of the toy model of \cite{mukamel-ssb-rw, mukamel-rev} in explaining the occurrence of spontaneous symmetry breaking in our model. Specifically, the model helps us to understand how, starting from the symmetric phase on the $\beta=0$ plane for $\gamma$ small ($< 1/2$), turning on arbitrarily small $\beta$ results in symmetry-broken phases. 
The toy model was initially devised to gain insight into the flipping process between the two states in the symmetry-broken phases of the original bridge model which has $\gamma=0$. The crucial step was to identify that, for small particle extraction rate $\beta$, typical configurations on a lattice of size $N$ are those composed of three blocks: a left block with an integer number $j$ of the negative particles, a right block with an integer number $k$ of the positive particles, and a central block with $N-j-k$ holes. Other configurations in which holes are present inside the particle segments do not play a role in determining the long-time behavior of the system and may be neglected. On the time scale set by the rate $\beta$, the dynamics of the system will involve transitions between various three-block configurations. 

In our model, the extraction rate equals $\beta+\gamma$. Now, consider the case of small $\beta+\gamma$, with $\beta \ll \gamma$. Typical configurations will again have three blocks, albeit with important differences from those in the original bridge model. On the time scale of $1/\beta$, most of the positive particles entering the lattice through exchanging the negative particles at the left end will have enough time to move away from the left end, so that the left block will have predominantly negative particles with the positive particles clustering towards the right end of the block. The crucial point to remember is that the left block will basically have no holes. Similarly, the right block will have predominantly positive particles towards the right end with some negative particles progressively clustered towards the left end of the right block. Again, the right block has basically no holes. All the holes are trapped between the left and the right blocks. Thus, the central block will have mostly holes with a few particles of both species here and there between the holes. A typical configuration will thus look like $----+----+++--++-++++0+00-0000+-+000-+---+--++-+--++++++$. The block of the negative particles at the extreme left end and the block of the positive particles at the extreme right end are long-lived in the limit of small $\beta+\gamma$. In the limit of small $\beta$, the dynamics of the system will involve transitions between such three-block configurations. 

Now that we have identified the left block of length $j$ with predominantly negative particles and the right block of length $k$ with predominantly positive particles, the arguments given in the context of the toy model equally apply to the problem at hand. Thus, the dynamics restricted to $(j,k)$ configurations will be that of a random walker in the first quadrant of the $(j, k)$ space within the triangle with corners at $(0,0),(N,0)$, and $(0,N)$. Starting with a system with only positive particles (i.e., from the point $(0,N)$), typical trajectories of the random walker will be biased towards the $k$ axis \cite{mukamel-ssb-rw}; see also \cite{mukamel-rev}. In order to flip to a system with only the negative particles, the trajectory will have to perform an atypical walk against the bias, starting from the point $(0,N)$ and ending on the $j$ axis without touching the $k$ axis. The probability of such a walk was calculated to be exponentially small in the system size \cite{mukamel-ssb-rw}. As a result, the flipping time $\tau(N)$ diverges exponentially with the system size, as confirmed by the results of our simulation (Fig. \ref{tauN vs. N}). Thus, the toy model of a biased random walker serves as a guide to understand intuitively the occurrence of spontaneous symmetry breaking in our model in the limit of small $\beta$ and $\gamma$, with $\beta \ll \gamma$.             
\section{Conclusions}
In this paper, we have revisited the problem of spontaneous symmetry breaking under nonequilibrium conditions in a one-dimensional model system with local dynamics and a finite state space for the local variables of the system. In this so-called bridge model, where two species of hard core particles are driven in opposite directions on an open lattice, earlier studies have shown the occurrence of SSB in the limit of small extraction rate $\beta$ of particles \cite{mukamel-ssb1, mukamel-ssb2}. In this study, we have allowed for boundary exchange of particles of one species into another with rate $\gamma$. This is the most general dynamical move at the boundaries which is consistent with the symmetry of the model and with the total asymmetry in the direction of motion of the particles. In this modified model, if the rate $\beta$ is zero, exact results predict only symmetric phases. On the other hand, if the rate $\beta$ is large, one expects, on physical grounds, only symmetric phases. Also, if the exchange rate $\gamma$ equals the particle injection rate $\alpha$, there is no SSB, as confirmed in \cite{schutz-pde}. Thus, one is left to wonder about the occurrence of SSB for small values of $\beta$ and non-zero $\gamma$. 

Here, we reported exact as well as mean-field results for the complete phase diagram of the modified bridge model in the $\alpha-\beta-\gamma$ space, showing regions for the symmetric and the symmetry-broken phases. Our results confirm the existence of SSB for non-zero $\beta$ and $\gamma$, provided both the rates are not too large, as have been quantified in the paper. Similar to the original bridge model, in the symmetry-broken phases, the system resides in one of two long-lived states, whose average lifetime grows exponentially with the system size. Our results are supported by extensive Monte Carlo simulations of the model. All these observations lead us to conclude that the symmetry breaking in the original bridge model is quite insensitive to additional dynamical moves, like those allowing for particle exchange at the boundaries, for a range of values of the exchange rate. It is left as an open problem to obtain the exact solution of the stationary state for the entire range of values of the parameters defining the model, for example, by including the boundary exchange rate $\gamma$ in the studies pursued in \cite{schutz-ssb1, schutz-ssb2}, and also, to study the model in higher dimensions.           
\label{conclusions}
\section{Acknowledgements}
This work was started during S. G.'s visit to the Weizmann Institute of Science (WIS) in May, 2008 while he was still a graduate student at the Tata Institute of Fundamental Research (TIFR), Mumbai, India. He gratefully acknowledges the WIS and the Sarojini Damodaran International Fellowship of TIFR for supporting his visit. This work was carried out in part while G. M. S. was Weston Visiting Professor at WIS. The support of the Israel Science Foundation (ISF) is gratefully acknowledged. 
\appendix
\section{Proof of non-existence of the power-law/low-density (pl/ld) phase within the mean-field approximation}
\label{app1}
The conditions for the existence of the pl/ld phase (with positive particles in the pl phase) are
\be
\alpha^+ \ge 1/2, ~~~~ \beta+\gamma \ge 1/2, ~~~~ \alpha^-<\beta+\gamma, ~~~~ \alpha^-<1/2.
\label{app1eq1}
\ee
Correspondingly, the particle currents satisfy $j^+=1/4$ and $j^{-}<1/4$. From Eq. (\ref{j-steady-lr}), it then follows that
\bea
&j^+=\alpha(1-p_1)-(\alpha-\gamma)m_1=(\beta+\gamma)p_N=1/4, \label{app1eq2} \\
\mathrm{and}& \nonumber \\
&j^-=\alpha(1-m_N)-(\alpha-\gamma)p_N=(\beta+\gamma)m_1 < 1/4. \label{app1eq3}
\eea

In order to prove the non-existence of the pl/ld phase, we will use the above expressions for the currents as well as the facts that the negative particles are in the low-density phase so that $0 < m_{N} < 1/2$ and that they have a flat density profile near the right boundary so that $j^{-}=m_{N}(1-m_{N})$. We get an expression for $m_{N}$ which in turn yields an expression for $m_{1}$. Now, since the positive particles are in the power-law phase, $p_{1} >1/2$, which, on using Eq. (\ref{app1eq2}), puts a bound on $m_{1}$. We will show that the derived expression for $m_{1}$ fails to satisfy this bound in the relevant parameter regime.

From Eq. (\ref{app1eq2}), we get 
\be
p_N=\frac{1}{4(\beta+\gamma)}.
\label{app1eq4}
\ee
Since the negative particles are in the low-density phase for which the density profile is flat at the entrance end, one has $j^{-}=m_{N}(1-m_{N})$. Then, from Eq. (\ref{app1eq3}), on using the expression for $p_N$ in Eq. (\ref{app1eq4}), we get
\be
m_N(1-m_N)=\alpha(1-m_N)-\frac{(\alpha-\gamma)}{4(\beta+\gamma)},
\label{app1eq5}
\ee
which gives a quadratic equation in $m_N$, and therefore,
\be
m_N=\frac{1+\alpha}{2}- \frac{1}{2}\sqrt{(1-\alpha)^2+\frac{(\alpha-\gamma)}{(\beta+\gamma)}}.
\label{app1eq6}
\ee
Here we have taken the negative root for $m_{N}$. This is because the negative particles are in the low-density phase, and hence, we should have $0< m_N <1/2$. The negative root satisfies these bounds, provided
\bea
&\alpha > \frac{(\alpha-\gamma)}{4(\beta+\gamma)}, \label{app1eq7a} \\
&(1-\alpha)^2+\frac{(\alpha-\gamma)}{(\beta+\gamma)} \ge 0, \label{app1eq7b} \\
\mathrm{and}& \nonumber \\
&\alpha < \sqrt{(1-\alpha)^2+\frac{(\alpha-\gamma)}{(\beta+\gamma)}}. \label{app1eq7c}
\eea
The inequality in Eq. (\ref{app1eq7c}) may be rearranged to give
\be
\beta < \frac{\alpha(1-2\gamma)}{2\alpha-1},
\label{app1eq8}
\ee
so that, in order to have a finite $\beta > 0$, we need to have either $\alpha > 1/2, ~~ 0 < \gamma <1/2$, or $0 < \alpha <1/2, ~~ \gamma >1/2$. Also, we have, from Eq. (\ref{app1eq1}),
\be
\beta+\gamma \ge \frac{1}{2}.
\label{app1eq9}
\ee
Let us introduce a new variable $x$ by the following equation,
\be
x=\frac{(\alpha-\gamma)}{(\beta+\gamma)}.
\label{xdefn}
\ee
For $x>0$ (when $\alpha > 1/2$ and $0<\gamma <1/2$), Eqs. (\ref{app1eq7a}), (\ref{app1eq7b}), (\ref{app1eq7c}) and (\ref{app1eq9}) give the following bounds on $x$, 
\be
2\alpha-1 < x \le 2\alpha.
\label{app1eq10}
\ee
On the other hand, for $x<0$ (when $0 < \alpha<1/2$ and $\gamma >1/2$), Eqs. (\ref{app1eq7a}), (\ref{app1eq7b}), (\ref{app1eq7c}) and (\ref{app1eq9}) allow for $x$ to lie in the following range:
\be
2\alpha-1 < x < 0.
\label{app1eq11}
\ee

Next, we derive an expression for $m_1$ from Eq. (\ref{app1eq3}) by substituting  $j^-=m_N(1-m_N)$, with $m_N$ given by Eq. (\ref{app1eq6}). We get 
\be
m_1=\frac{1}{4(\beta+\gamma)}\left[1-\left(\alpha-\sqrt{(1-\alpha)^2+\frac{(\alpha-\gamma)}{(\beta+\gamma)}}\right)^2\right].
\label{app1eq12}
\ee
On the other hand, Eq. (\ref{app1eq2}) yields
\be
m_1=\frac{\alpha}{(\alpha-\gamma)}(1-p_1)-\frac{1}{4(\alpha-\gamma)}.
\label{app1eq13}
\ee
Since the positive particles are in the power-law phase, we have $p_1 > 1/2$, i.e., $1-p_1 < 1/2$. Combined with the last equation, this gives 
\bea
&m_{1} < \frac{2\alpha-1}{4(\alpha-\gamma)} \mathrm{~for~} \alpha > 1/2 \mathrm{~and~} \gamma < 1/2, \label{app1eq13a} \\
\mathrm{and}& \nonumber \\
&m_{1} > \frac{1-2\alpha}{4(\gamma-\alpha)} \mathrm{~for~} \alpha < 1/2 \mathrm{~and~} \gamma > 1/2, \label{app1eq13b}
\eea
with $m_{1}$ given in Eq. (\ref{app1eq12}).
In terms of the variable $x$, it follows from the above inequalities that, for positive $x$ with $2\alpha-1 < x \le 2\alpha$, we must have
\be
x\left[1-\left(\alpha-\sqrt{(1-\alpha)^2+x}\right)^2\right]-(2\alpha-1) < 0, 
\label{app1eq14}
\ee
while, for negative $x$, the above inequality has to be satisfied for $2\alpha-1 < x < 0$. It can be checked that in either case, the inequality has no solution, implying that the pl/ld phase does not exist in our model within the mean-field theory.\\


\begin{thebibliography}{99}
\bibitem{mukamel-rev}Mukamel D 2000 \textit{Soft and Fragile Matter: Nonequilibrium Dynamics, Metastability and Flow} ed M E Cates and M R Evans (Institute of Physics Publishing, Bristol); also, e-print:arXiv:cond-mat/0003424.  

\bibitem{schutz-twospecies}Sch\"{u}tz G M 2003 \textit{J. Phys. A} \textbf{36} R339.

\bibitem{mukamel-ssb1}Evans M R, Foster D P, Godr\`{e}che C and Mukamel D 1995 \textit{Phys. Rev. Lett.} \textbf{74} 208.

\bibitem{mukamel-ssb2}Evans M R, Foster D P, Godr\`{e}che C and Mukamel D 1995 \textit{J. Stat. Phys.} \textbf{80} 69.

\bibitem{mukamel-ssb-rw}Godr\`{e}che C, Luck J M, Evans M R, Mukamel D, Sandow S and Speer E R 1995 \textit{J. Phys. A} \textbf{28} 6039.

\bibitem{schutz-ssb1}Willmann R D, Sch\"{u}tz G M and Gro{\ss}kinsky S 2005 \textit{Europhys. Lett.} \textbf{71} 542.

\bibitem{schutz-ssb2}Gro{\ss}kinsky S, Sch\"{u}tz G M and Willmann R D 2007 \textit{J. Stat. Phys.} \textbf{128} 587.

\bibitem{popkov-ssb}Popkov V and Peschel I 2001 \textit{Phys. Rev. E} \textbf{64} 026126.

\bibitem{levine-ssb}Levine E and Willmann R D 2004 \textit{J. Phys. A} \textbf{37} 3333.

\bibitem{lipowsky-ssb}Klumpp S and Lipowsky R 2004 \textit{Europhys. Lett.} \textbf{66} 90.

\bibitem{kolomeisky}Pronina E and Kolomeisky A B 2007 \textit{J. Phys. A} \textbf{40} 2275.

\bibitem{jiang-1}Jiang R, Wang R, Hu M, Jia B and Wu Q 2007 \textit{J. Phys. A} \textbf{40} 9213.

\bibitem{jiang-2}Jiang R, Hu M, Jia B, Wang R and Wu Q 2007 \textit{Phys. Rev. E} \textbf{76} 036116.

\bibitem{jiang-3}Yuan Y, Jiang R, Wang R, Wu Q and Zhang J 2008 \textit{J. Phys. A} \textbf{41} 035003.

\bibitem{bridge-junction}Popkov V, Evans M R and Mukamel D 2008 \textit{J. Phys. A} \textbf{41} 432002. 

\bibitem{asep-rev}Sch\"{u}tz G M 2001 \textit{Phase Transitions and Critical Phenomena} ed C. Domb and J. L. Lebowitz (Academic, London).

\bibitem{schutz-pde}Popkov V and Sch\"{u}tz G M 2004 \textit{J. Stat. Mech.: Theory Exp.} P12004.

\bibitem{second-class}Ferrari P, Kipnis C and Saada E 1991 \textit{Ann. Prob.} \textbf{19} 226.

\bibitem{asep-mft}Derrida B, Domany E and Mukamel D 1992 \textit{J. Stat. Phys.} \textbf{69} 667.

\bibitem{asep-exact1}Derrida B, Evans M R, Hakim V and Pasquier V 1993 \textit{J. Phys. A} \textbf{26} 1493.

\bibitem{asep-exact2}Sch\"{u}tz G M and Domany E 1993 \textit{J. Stat. Phys.} \textbf{72} 277.

\bibitem{asep-shock}Kolomeisky A B, Sch\"{u}tz G M, Kolomeisky E B and Straley J P 1998 \textit{J. Phys. A} \textbf{31}, 6911.

\bibitem{ahr-ssb}Arndt P F, Heinzel T and Rittenberg V 1998 \textit{J. Stat. Phys.} \textbf{90} 783.

\bibitem{clincy-ssb}Clincy M, Evans M R and Mukamel D 2001 \textit{J. Phys. A} \textbf{34}, 9923.

\bibitem{zia-ssb}Erickson D W, Pruessner G, Schmittmann B and Zia R K P 2005 \textit{J. Phys. A} \textbf{38} L659. 



\end{thebibliography}
\end{document}